\documentclass[12pt,aps,showpacs,eqsecnum,nofootinbib,floatfix]{revtex4}
\usepackage{CJK}
\usepackage{bbding}
\usepackage{pifont}
\usepackage{subfigure}
\usepackage{epsfig}
\usepackage{array}
\usepackage{amsmath}
\usepackage{amsfonts}
\usepackage{amssymb}
\usepackage{color}
\usepackage{graphicx}
\usepackage{mathrsfs}
\usepackage{multirow}
\usepackage{caption}

\def\be{\begin{equation}}
\def\ee{\end{equation}}
\def\bea{\begin{eqnarray}}
\def\eea{\end{eqnarray}}

\newcommand{\omits}[1]{}

\begin{document}

\begin{CJK*}{GBK}{kai}

\title{microstructure and continuous phase transition of RN-AdS black hole}
\author{Xiong-Ying Guo$^{a,b}$, Huai-Fan Li$^{a,b}$\footnote{Email: huaifan.li@stu.xjtu.edu.cn; huaifan999@sxdtdx.edu.cn(H.-F. Li)}, Li-Chun Zhang$^{a,b}$, Ren Zhao$^{b}$}

\medskip

\affiliation{\footnotesize$^a$ Department of Physics, Shanxi Datong
University,  Datong 037009, China\\
\footnotesize$^b$ Institute of Theoretical Physics, Shanxi Datong
University, Datong 037009, China}

\begin{abstract}
As is well known that RN-AdS black hole has a phase transition which is similar to that of van der Waals system. The phase transition depends on the electric potential of the black hole
and is not the one between a large black hole and a small black hole. On this basis, we introduce a new order
parameter and use the Landau continuous phase transition theory to discuss the critical phenomenon of RN-AdS black hole
and give the critical exponent. By constructing the binary fluid model of black hole molecules, we investigate the microstructure of black holes.
Furthermore, by studying the effect of the spacetime scalar curvature on the phase transition, we find that the charged and uncharged molecules of black holes are with different microstructure red which is like fermion gas and boson gas.
\end{abstract}

 \maketitle

\section{Introduction}
As an interdisciplinary area of general relativity, quantum mechanics, thermodynamics and statistical
physics, particle physics and string theory, black hole physics plays a very important role in modern physics.
The investigation of the thermal properties and the internal microstructure of black holes has always been one of the topics of interest to theoretical physicists. In recent years, the first law of black hole thermodynamics has been obtained by matching the cosmological constant in a AdS spacetime. Then the discussion of the thermodynamic properties of AdS and dS black holes by comparing the black hole state parameters
with the van de walls(vdW) equation has been widely concerned~\cite{Mann1207,Mann1301,Kubiznak1608,Sekiwa0602,Cai1306,Hennigar2017,Hendi1702,Hendi1803,Hendi2017,Dayyani1709,
Ovgun1710,Ma2018,Javed1805,Zhao1405,Ma1604,Ma1707,Hendi1612,Dayyani2017,Zou1702,Zou2017,
Cheng1603,Mann1610,Banerjee1611,Banerjee1109,Banerjee1203,Bhattacharya2017,Zeng2017,Dehyadegari,Cai1606,Zhang1502,Jafarzade1803}.

It is found that the second-order phase transition point of the black hole is independent of the conjugate variables~\cite{Zhao1405,Ma1604,Ma1707,Hendi1612,Dayyani2017,Zou1702,Zou2017,
Cheng1603,Mann1610,Banerjee1611,Banerjee1109,Banerjee1203,Bhattacharya2017}.
In the case of vdW system with the fixed temperature and pressure, the microstructure and the physical properties of the system changes when it experiences the liquid-gas phase transition. For the research of black hole system, it is known that when the system is in isothermal or isobaric processes, the $P-V$ curve or the $T-S$ curve of system is discontinuous. This result meet the requirements of thermodynamic equilibrium stability. we known that the system has latent heat of phase transition when it passes through the two-phase coexistence curve. Since the thermal system has different physical properties in the two phases, thus the system has different microstructure in different phases. Research on the microstructure of black holes is the foundation of the theory of quantum gravity and the bridge between quantum mechanics and the gravity theory. The study of the classical thermodynamic properties and microstructure of black holes will provide an important window for the exploration of quantum gravity. It is hoped that the knowledge of the microstructure of black holes can be revealed by studying the phase transition of black holes.

Recently based on the study of the thermodynamic properties of black holes, some researchers have adopted Boltzmann's profound idea, i.e., a black hole can change its temperature by
absorbing and emitting matter. An object with temperature has a microstructure. The authors have shown that black holes are also composed of effective molecules, which carry the microscopic degrees of freedom of black hole entropy ~\cite{Wei2015}. Based on this hypothesis, the concept of molecular density of black holes is introduced to obtain the phase transition of black holes. And it is given by
\begin{equation}
\label{eq1}
m=\frac{1}{v} = \frac{1}{2l_p^2 r_ + },
\end{equation}
where $l_p $ is the Planck length, $l_p = \sqrt {\hbar G / c^3}$.
Introduced $m\equiv (m_{SBH} - m_{LBH} ) / m_c $ with the aid of the
critical number density $m_c = 1 / (2\sqrt 6Q)$. According to the research, the order parameter has a non-zero value when the system passes through the first-order phase transition of the small black hole and the large one. The value of order parameter gradually decrease as the temperature increases, and it equates to zero at the critical point. To some extent, the microstructure of small black holes and large black holes will converge at the critical point. It is found that the variation of the molecular density of the black hole is the internal cause of the black hole phase transition~\cite{Wei2015,Blanchet,Miao1712,Miao1711}. From the Eq.(\ref{eq1}), it seems
that the molecular density $m$ is only the function of black hole event horizon $r_+$, while it is related with other parameters of black holes such as the electric charge and magnetic charge.

Recent studies have shown that the RN-AdS black hole has the same thermodynamic characteristics as that for vdW system. And the cosmological constant is interpreted as the pressure in the thermodynamic system. With these issues, the thermodynamic characteristics and critical phenomena of AdS and dS black holes are studied, and the first-order and second-order phase transitions of RN-AdS black holes with vdW system are obtained. Furthermore, the critical boundary point of phase transition has the same critical exponent and scale rate~\cite{Mann1207,Wei2015,Miao1610,Wei1209}. However, Schwarzschild de Sitter(SdS) black holes without charge has no phase transition like vdW system~\cite{Zhang1409}. Therefore, the charge of the black hole plays a key role in the phase transition, and the effect of the black hole charge on the phase transition must be considered in the microstructure theory of the black hole.

In this paper, based on the study of the black hole phase transition with different conjugate variables, we find that the phase transition of RN-AdS black hole is not just the phase transition of small black hole and large black hole, but it is determined by the potential of the black hole. The phase of the black hole is in the high potential phase, the low potential phase and the medium potential phase. The neutral potential corresponds to the liquid-vapor coexistence region of the vdW system. As RN-AdS black hole molecules are subject to electric potential (electric field), the black hole molecules generate orientation polarization and displacement polarization, so that the black hole molecules have a certain orientation under the dual effects of different electric potential and thermal motion, the black hole molecules' orientation is determined by the black hole's phase. Under the guidance of this idea, in the second part, we select the electric potential as the order parameter to study the black hole phase transition, and analyze the continuous phase transition of RN-AdS black hole using the Landau continuous phase transition theory. The microscopic interpretation and critical exponents of the black hole phase transition are given. As a unique perspective, thermodynamic geometry plays an important role in the study of black hole phase transition. Next, the properties of the microstructure of black hole molecules are investigated by analyzing the influence of the charge of the black hole molecules on the positive and negative values of the $R-\chi$ curve. Finally, we give the summary. For simplicity, in this paper we adopt the following the units $\hbar = c = k_B = G = 1$ .

\section{Phase transition of RN-AdS black hole}
\label{re}
We find that the phase transition position of RN-AdS black hole is
invariable with different conjugate variables $(P - V)$ and $(T -S)$, and the thermodynamic properties is invariable with different
conjugate variables. As is well known we study the molar mass system with the study of the phase transition of thermodynamics system, such as
the phase transition for molar vdW system. Keeping the charge $Q$ of system is constant, we know that the charge $Q$ of RN-AdS
black hole is analogous to the molar mass of vdW system in studying RN-AdS black hole phase transition. Taking the conjugate variables $(P - V)$
and the charge $Q$ of RN-AdS black hole as a constant. The horizontal axis of the two phase coexistence region is $V_2$ and $V_1$, respectively, the vertical axis is $P_0$,
which depends on the radius $r_+$ of the black hole event horizon. According to the condition of equilibrium stability,
we can obtain the following expression by the Maxwell's equal area laws,
\begin{equation}
\label{eq2}
P_0 (V_2 - V_1 ) = \int\limits_{V_1 }^{V_2 } {PdV}.
\end{equation}
Taking the conjugate variables $(T -S)$ and the invariable cosmological constant $l$, the horizontal axis of the two phase coexistence region is $S_2$ and $S_1$, respectively. The vertical axis is $T_0,(T_0 \le T_c )$
which depends on the radius $r_+$ of the black hole event horizon. According to the condition of equilibrium stability,
we can obtain by the Maxwell's equal area laws,
\begin{equation}
\label{eq3}
T_0 (S_2 - S_1 ) = \int\limits_{S_1 }^{S_2 } {TdS_ + } = \int\limits_{r_1
}^{r_2 } {\frac{1}{2}\left( {1 + \frac{3r_ + ^2 }{l^2} - \frac{Q^2}{r_ + ^2
}} \right)dr_ + }.
\end{equation}
From Eqs.(\ref{eq2}) and (\ref{eq3}), we can obtain
\begin{equation}
\label{eq4}
\chi x^2f_Q^{3 / 2} (x)\frac{1}{3\sqrt 6 } = (1 + x)
\end{equation}
\begin{equation}
\label{eq5}
r_2^2 = \frac{Q^2}{x^2}\frac{(1 + 2x - 6x^2 + 2x^3 + x^4)}{(1 - x)^2} =
\frac{Q^2}{x^2}(1 + 4x + x^2) = Q^2f_Q (x).
\end{equation}
where $x = r_1 / r_2 $, $r_1$ and $r_2 $ is the black hole event horizon radius
for a  first-order phase transition point, respectively. The temperature $T_0=\chi T_c$, here $ T_c $ is critical temperature,  $0 <\chi\le 1$ is a constant.
When $\chi$ is determined, we can obtain
the value of $x$ and $r_2 $ from Eqs. (\ref{eq4}) and (\ref{eq5}). When $x =
1$, we obtain the critical values from (\ref{eq4}) and (\ref{eq5}).

The lowest point of the $P-V$ graph satistfies $P = 0$, $\left(
{\frac{\partial P}{\partial V}} \right)_T = 0$, the corresponding temperature $T_0 =
\frac{1}{6\sqrt 3 \pi Q}$, that is,  the minimum value of $\chi
$ is $\chi = \frac{1}{\sqrt 2 }$. So Eq.(\ref{eq4}) can be written as
\begin{equation}
\label{eq6}
x^2f_Q^{3 / 2} (x)\frac{1}{6\sqrt 3 } = (1 + x).
\end{equation}
We can obtain the solution $x=x_0$ for equation
\begin{equation}
\label{eq7}
1 + 12x - 57x^2 - 128x^3 - 57x^4 + 12x^5 + x^6 = 0.
\end{equation}
From Eq.(\ref{eq5}), we find that there is a sudden change in the potential of the black hole molecules
when the phase transition of black hole happens with the temperature below $T_0$.
This also reflects the fact that the microstructure of black hole molecules in different
phases is inconsistent.When the temperature is determined, the two phase black hole molecules
are respectively at the electric potential
\begin{equation}
\label{eq8}
\varphi _2 = \frac{Q}{r_2 } = \frac{1}{f_Q^{1 / 2} (x)},
\quad
\varphi _1 = \frac{Q}{r_1 } = \frac{1}{xf_Q^{1 / 2} (x)} \quad .
\end{equation}
where $x$ is given by Eq.(\ref{eq4}). We take the order parameter
\begin{equation}
\label{eq9}
\varphi (T) = \frac{\varphi _1 - \varphi _2 }{\varphi _c } = \frac{r_c (1 -
x)}{r_2 x} = \frac{\sqrt 6 (1 - x)}{f_Q^{1 / 2} (x)x}
\end{equation}
\begin{figure}[!htbp]
\center{\includegraphics[width=7cm,keepaspectratio]{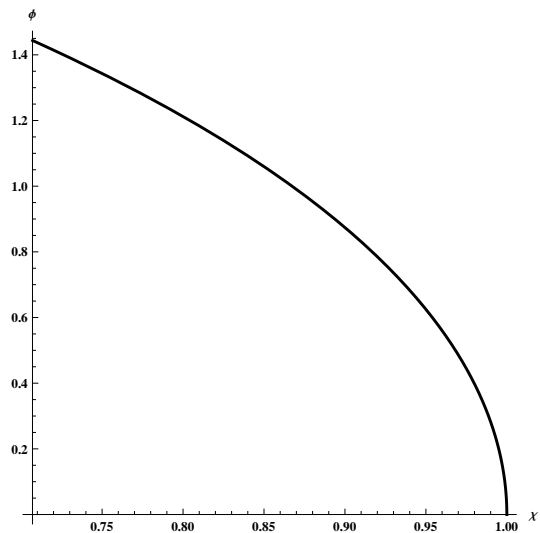}\hspace{0.5cm}}\\
\captionsetup{font={scriptsize}}
\caption{(Color online) $\varphi (T) - \chi $\label{PVR}}
\end{figure}
Recently the microstructure of the phase transition of black holes is studied
~\cite{Wei2015,Blanchet,Miao1712,Miao1711}.
It is found that the phase transition between the size of the black hole is due to the
different density of molecules inside the black hole.The effect of the black hole charge on the phase
transition is not shown in the calculation. Now let's reexamine the physical mechanism of a RN-AdS
black hole when we consider the effect of charge on the phase transition.

The Landau's continuous phase transition theory is characterized by the change
of the degree of material order and the change of the symmetries accompanying
it,So the phase transition of a charged black hole is also a change of symmetry.
When the temperature of the black hole is lower than the critical point, the
black hole molecules are in the phase 1 with high potential. The black hole
molecules in the black hole are subject to the action of strong electric
potential (electric field) to generate a certain orientation and displacement
polarization, which has a certain orientation. When black hole molecules are
in a relatively ordered state with low symmetry, they are in phase 2
with low potential at the same temperature, the electric potential (electric field)
that causes the black hole molecules to produce orientation decreases.
The order degree of black hole molecules is relatively low, but its symmetry is
relatively high. As the temperature increases, the thermal motion of
the black hole molecules tends to weaken the sequence orientation.
But just the temperature is not very high, there's still a certain orientation of black
hole molecules, and that's why the orientation of black hole molecules decreases with temperature.
When the black hole temperature is above the critical temperature, the thermal motion of
the black hole molecules increases, causing the black hole molecules to approach zero.
The phase below critical temperature with lower symmetry and higher order, order parameter
$\varphi(T)$ is non-zero. the phase above critical temperature with higher symmetry and lower order,
order parameter $\varphi(T)$ is zero. As the temperature decreases, the order parameter
$\varphi (T)$ changes continuously from zero to non-zero at the critical point.

Landau believed that the order parameter $\varphi(T)$ near the critical point $T_c$ is
a small amount, thus the Gibbs function $G(T,\varphi)$ can be expanded to the power of $\varphi(T)$ near $T_c$.
Considering spacetime, the transformation of $Q \leftrightarrows -Q$ is symmetric. When the electric potential is
used as the order parameter and the thermodynamic quantity of the black hole is expanded in
order parameter perturbation series. The odd order terms of the series are absent. So
\begin{equation}
\label{eq10}
G(T,\varphi ) = G_0 (T) + \frac{1}{2}a(T)\varphi ^2 + \frac{1}{4}b(T)\varphi
^4 + \cdots ,
\end{equation}
here $G_0 (T)$ is the Gibbs function at $\varphi (T) = 0$. We can determine the
function relation between $\varphi (T)=0$ and temperature $T$ with the minimum condition of stable
equilibrium Gibbs function at $T,P$ is invariant. Note that we write the Gibbs function as $G(T,\varphi)$,
where $\varphi$ is not independent variable. In a stable equilibrium state, $G(T)$ should have a minimum value
\begin{equation}
\label{eq11}
\frac{\partial G}{\partial \varphi } = \varphi (a + b\varphi ^2) = 0,
\end{equation}
\begin{equation}
\label{eq12}
\frac{\partial ^2G}{\partial \varphi ^2} = a + 3b\varphi ^2 > 0,
\end{equation}
Eq. (\ref{eq11}) have three solutions
\begin{equation}
\label{eq13}
\varphi = 0,
\quad
\varphi = \pm \sqrt { - \frac{a}{b}} ,
\end{equation}
This solution $\varphi=0$ represents a disordered state, that corresponds to the temperature range of $T > T_c $.
Substituting $\varphi = \pm \sqrt { - \frac{a}{b}}$ into (\ref{eq11}), we find that $a<0$ at $T<T_c$ and $a>0$ at
$T>T_c$. This nonzero solution $\varphi = \pm \sqrt { - \frac{a}{b}}$ represents an ordered state, that corresponds to the temperature range of $T < T_c $.  The order parameters continuously change from
zero to non-zero at $T_c$, so there should be $a = 0$ at $T = T_c$. In the critical neighborhood we can simply take
\begin{equation}
\label{eq14}
a = a_0 \left( {\frac{T - T_c }{T_c }} \right) = a_0 t,
\quad
a_0 > 0,
\end{equation}
and
\begin{equation}
\label{eq15}
b(T) = b,
\end{equation}
$b$ is a constant. $\varphi = \pm \sqrt { - \frac{a}{b}} $ is real numbers and $a <
0$ at $T < T_c $, so the constant $b > 0$. We can obtain
$$
\varphi = 0,
\quad
t > 0
$$
\begin{equation}
\label{eq16}
\varphi = \pm \left( {\frac{a_0 }{b}} \right)^{1 / 2}( - t)^{1 / 2},
\quad
t < 0
\end{equation}
here the critical exponent $\beta = 1 / 2$. Substituting Eq. (\ref{eq15}) into Eq.(\ref{eq9}), we get
\[
G(T,\varphi ) = G_0 (T),
\quad
T > T_c
\]
\begin{equation}
\label{eq17}
G(T,\varphi ) = G_0 (T) - \frac{a_0^2 }{4b}\left( {\frac{T - T_c }{T_c }}
\right)^2,
\quad
T < T_c .
\end{equation}
From Eq. (\ref{eq16}) we get that the heat capacity at the critical point is jumping. For $C = - T\left( {\frac{\partial
^2G}{\partial T^2}} \right)$, we can get
\begin{equation}
\label{eq18}
C(t < 0)_{t = 0} - C(t > 0)_{t = 0} = \frac{a_0^2 }{2bT_c },
\quad
t < 0.
\end{equation}
So the jump curve of heat capacity is $lambda$ shape. Eq. (\ref{eq17}) indicates that the heat capacity of ordered phase is greater than that of disordered phase, and the sudden change of heat capacity at $t= 0$ is limited. So we know that the critical parameter $\alpha = \alpha ' = 0$.
When the pressure change is negligible, the total differential of Gibbs free energy as temperature $T$ and electric potential $\varphi$ is given~\cite{Wei1209}
\begin{equation}
\label{eq19}
dG = - SdT - Qd\varphi
\end{equation}
For Eq.(\ref{eq19}), we get
\begin{equation}
\label{eq20}
 - Q = \left( {\frac{\partial G}{\partial \varphi }} \right)_T = a\varphi +
b\varphi ^3,
\end{equation}
So
\begin{equation}
\label{eq21}
 - \left( {\frac{\partial \varphi }{\partial Q}} \right)_T = \frac{1}{a +
3b\varphi ^2} = \left\{ {{\begin{array}{*{20}c}
 {\frac{1}{a_0 }t^{ - 1},} \hfill & {t > 0} \hfill \\
 {\frac{1}{2a_0 }( - t)^{ - 1},} \hfill & {t < 0} \hfill \\
\end{array} }} \right.,
\end{equation}
From Eq.(\ref{eq19}) we obtain $_{ }\gamma = \gamma'=1$. When $T = T_c $ and $a =
0$, we can obtain the relation of charge $Q$ and electric potential $\varphi $ with Eq.(\ref{eq19})
\begin{equation}
\label{eq22}
Q \propto \varphi ^3.
\end{equation}
The critical exponent $\delta = 3$. The critical exponent is same as that in Refs.~\cite{Mann1207,Dehyadegari,Wei1209}.
From Eq. (2.17), we can get
\begin{equation}
\label{eq23}
S = - \left( {\frac{\partial G}{\partial T}} \right)
\end{equation}
Substituting (\ref{eq19}) into (\ref{eq22}), we can obtain the entropy with disordered phase is $S = S_0 $,
and the entropy with ordered phases is $S = S_0 + \frac{a_0^2 t}{2bT_c }$.
When $t = 0$, the ordered phase entropy is equal to the disordered phase entropy, that is,
the black hole entropy is continuous at the critical point.

\section{Thermodynamics geometry}
\label{eff}

From the discussion in part \ref{re}, we find that Eq.(\ref{eq9}) contain parameters of $a$ and $b$ related
to specific system characteristics, but the critical exponent given above is irrelevant to $a$ and $b$.
And we know that there are similar problems in the general thermodynamic system. The radical cause is that when we
discuss the continuous phase transition, we do not consider the strong fluctuation of order parameters in
the critical point domain. Fortunately, the famous Ruppeiner geometry comes from the thermodynamic fluctuation theory.
By studying the singularity of the curvature scalar corresponding to thermodynamic geometry,
the phase change structure of black hole is revealed~\cite{Ruppeiner2008,Ruppeiner1995}.
Therefore, we can reveal the microstructure of black hole molecules by studying the Ruppeiner geometry, for RN-AdS
black hole which has scalar curvature~\cite{Blanchet}.
\begin{equation}
\label{eq24}
R = \frac{2\pi Q^2 - S}{8PS^3 + S^2 - \pi SQ^2}.
\end{equation}
Furthermore, $R$ can be rewritten in terms of electric potential $n^2 = \frac{Q^2}{r_
+ ^2 }$ as
\begin{equation}
\label{eq25}
R = \frac{((n / n_c )^2 - 3)}{3\pi Q^2\left( {3(P / P_c )(n_c / n)^2 + 6 -
(n / n_c )^2} \right)}\left( {\frac{n}{n_c }} \right)^2.
\end{equation}
At the same temperature $R$ is divided into two, a representative
$n^2 =\varphi_2^2$, another representative $n^2=\varphi_1^2$.
Substitute Eq.(\ref{eq8}) into Eq.(\ref{eq25}) and get
\begin{equation}
\label{eq26}
R_2 = \frac{x^2(2 - f_Q (x))}{\pi Q^2f_Q (x)\left( {3 + x^2f_Q (x) - x^2)}
\right)} = - \frac{x^2(1 + 4x - x^2)}{4\pi Q^2(1 + 4x + x^2)(1 + x)},
\end{equation}
\begin{equation}
\label{eq27}
R_1 = \frac{(2 - x^2f_Q (x))}{\pi Q^2x^2f_Q (x)\left( {3x^2 + x^2f_Q (x) -
1} \right)} = \frac{1 - 4x - x^2}{4\pi Q^2x(1 + 4x + x^2)(1 + x)}.
\end{equation}
here $P/P_c = \frac{3\times 96\pi Q^2}{8\pi r_2^2 (1 + 4x + x^2)} =
\frac{36x^2}{(1 + 4x + x^2)^2}$, $P_c = \frac{1}{96\pi Q^2}$.
From Eqs. (\ref{eq26}) and (\ref{eq27}), we can plot the curve of $R - \chi$($x_0 < x \le 1)$~\cite{Zangeneh1602},
\begin{figure}[!htbp]
\center{\includegraphics[width=7cm,keepaspectratio]{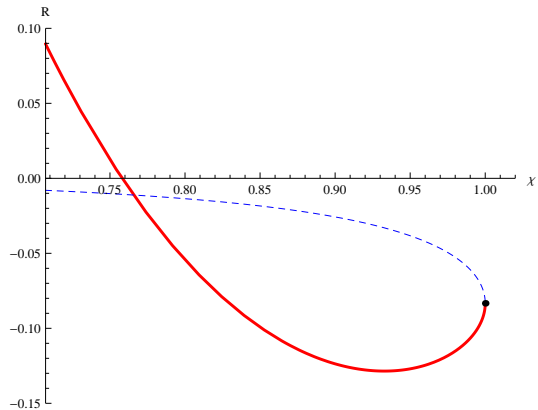}\hspace{0.5cm}}\\
\captionsetup{font={scriptsize}}
\caption{(Color online) $R -\chi$\label{PVR}}
\end{figure}

RN-AdS black holes are charged black holes, there are two types of black hole molecules,
a class of charged black hole molecules and a class of uncharged black hole molecules that
themselves carry microscopic degrees of freedom of black hole entropy. In order to discuss the role of
charged black hole molecules in the phase transition, we establish the black hole two fluid model that
charged black hole is a mixture of charged black hole molecules and uncharged black hole molecules.

Suppose the degree of freedom of the charged black hole molecule is $N_q$, and the
total molecular degree of freedom of the black hole is $N$.
According to the degree of freedom ratio of charged and uncharged molecules
in a black hole, it can be divided into three regions: $N_q / N
= n_0, N_q / N > n_0 $ and $N_q / N < n_0 $. When $N_q / N = n_0$, we think the degree of freedom of black
hole molecules is moderate; When $N_q / N > n_0 $, the black hole molecules are arranged in an orderly
way and have a low degree of freedom with the action of electric potential; When $N_q / N < n_0$,
black hole molecules are disordered and the degree of freedom of black hole molecules is relatively
high with the action of electric potential.

In the $R-\chi$ curve, the electric potential of $R_2 $ is lower than that of $R_1 $ at the same temperature, that is $\varphi _2^2 <
\varphi _1^2 $, and this curve $N_q / N < n_0 $ indicates $R_2 < 0$. So $R_2
$ corresponds to the fact that the black hole molecules are in a disordered state where the symmetry of the black hole molecules is high.
We discuss the $R_1-T$ curve in three parts:
\begin{itemize}
\item The phase of $R_1<0$ is in the region $\sqrt 5 - 2 < x \le 1$.
The potential $\varphi _1^2 $ in the region of $R_1$ corresponds to the potential $\varphi _2^2
$ in $R_2 $ at the same temperature, with $\varphi _2^2 < \varphi _1^2 $, and we have $N_q / N < n_0$ in the region.

\item The phase of $R_1>0$ is in the region $x_0 \le x < \sqrt 5 - 2$, here corresponding temperature of $x_0$
is $T = \frac{T_c}{\ SQRT 2}$. The potential $\varphi _1^2 $ in the region of $R_1$ corresponds to the potential $\varphi _2^2
$ in $R_2$ at the same temperature, with $\varphi _2^2 < \varphi _1^2 $, and we have $N_q / N > n_0$ in the region.

\item The phase of $R_1=0$ is at the point $2 - x^2f_Q (x) = 0x = \sqrt 5 -
2$, here the corresponding temperature $T = \frac{3\sqrt 3 }{2}(7 - 3\sqrt 5 )T_c $ from (\ref{eq4}), $R_1 =
0$ that correspond to the potential $\varphi _1^2 =\frac{1}{2}$.
\end{itemize}
According to the relation between the degree
of molecular freedom of the black hole and the horizon area of the black hole~\cite{Wei2015,Ruppeiner2008,Altamirano2014}
\begin{equation}
\label{eq28}
N = \frac{A}{\gamma l_p^2 },
\quad
n = \frac{N}{V} = \frac{3}{\gamma l_p^2 r_ + },
\end{equation}
we get
\begin{equation}
\label{eq29}
\varphi _1^2 = \frac{Q^2}{r_1^2 } = \frac{N_q }{N} = n_0 = \frac{1}{2}.
\end{equation}
The electric potential of $R_2$ at the same temperature $\varphi _2^2 = \frac{9 - 4\sqrt 5 }{2}$, we get $\varphi
_2^2 < \varphi _1^2 $.

In sum, we get that the phase of $R_1 >0$ black hole has a higher degree of order and a lower degree of symmetry;
In the phase of $R_1 = 0$, the degree of order of the black hole molecules is moderate and the degree of symmetry is moderate;
in the phase of $R_1 <0 $, the degree of order of the black hole molecules is lower and the degree of symmetry is higher.
When $x = 1$, $R_1 = R_2$, and $\varphi _1^2 =
\varphi _2^2 = \frac{N_q }{N} =
\frac{1}{6}$. That is, when the ratio of the free degree of charged molecules in the black hole molecules is $1 /
6$,the corresponding critical point is the demarcation point of order and disorder. When $\chi = \frac{1}{\sqrt 2 }$ and $x = x_0
$, the ratio of the degree of freedom of charged molecules to the total degree of freedom is $\frac{1}{1
+ 4x_0 + x_0^2 }$, where $x_0 $ is given by Eq.(\ref{eq6}).

\section{discussions and conclusions}
\label{con}
Since the RN-AdS black hole has a phase transition similar to the vdW system,
it can be assumed that they may have a similar microstructure.
At the microscopic level the fluid is made up of individual molecules. Therefore there is the assumption that
the black hole is composed of the black hole molecules and these molecules carry microscopic degrees of freedom of the black hole entropy ~\cite{Wei2015,Blanchet,Miao1712}.
For the RN-AdS black hole, the phase transition at a
certain temperature is not the one from a large black hole to
a small black hole, but is determined by electrical potential. The charges $Q$ of black hole play a key role in phase transition. It is known from Eq.(\ref{eq24}) that $R<0$ for $Q\rightarrow0$.
However, all the black hole molecules of SAdS black hole are composed of uncharged molecules. According to
the statistical interpretation of the scalar curvature $R$ symbol, we known that the scalar curvature is negative for the black hole system with
the uncharged black hole molecules, while it is positive for that of charged black hole molecules.
 B.Mirza et.al argue that the scalar curvature
is $R >0$ for boson gas and it is $R<0$ for fermion gas~\cite{Mirza2008,Mirza2009,Zangeneh1602,Mirza2013}. we find that the charged and uncharged molecules of black holes are with different microstructure which is like fermion gas and boson gas.

On the other hand, for anyon gas scalar curvature $R > $0 (or $R <0$) means that the average interaction between particles is repulsive or attractive, while the average interaction is zero as $R =0$~\cite{Mirza2009}. We can also think of black hole molecules as similar to anyon gas. Therefore, the three different regions of scalar curvature $R$ value of black hole system can
also be understood as that the average interaction of black hole molecules has repulsive, attraction and no interaction respectively~\cite{Wei2015,Miao1712,Miao1711}.

This work reveals the microscopic structure of the black hole from another view. This work can be
extended to higher dimensional and rotating AdS black hole. These conclusions help explore the
microstructure of black holes.

\section*{Acknowledgements}
We would like to thank Prof. Zong-Hong Zhu and Meng-Sen Ma for their indispensable discussions and comments. This work was supported by the Young Scientists Fund of the National Natural Science Foundation of China (Grant No.11205097), in part by the National Natural Science Foundation of China (Grant No.11475108), Supported by Program for the Innovative Talents of Higher Learning Institutions of Shanxi, the Natural Science Foundation of Shanxi Province,China(Grant No.201601D102004) and the Natural Science Foundation for Young Scientists of Shanxi Province,China (Grant No.2012021003-4).

\end{CJK*}

\end{document}